\documentclass[twocolumn, floatfix,showpacs,  prb]{revtex4}
\usepackage{amsfonts, url, amsmath}
\usepackage[dvips]{graphicx}

\begin{document}

\title{\bf  Reconstruction of silicon surfaces: a stochastic optimization
problem}

\author{Cristian V. Ciobanu$^{1}$ and Cristian Predescu$^{2}$}

\affiliation{$^{1}$Division of Engineering, Brown University,
Providence, RI 02912 }

\affiliation{$^{2}$Department of Chemistry and Kenneth S. Pitzer
Center for Theoretical Chemistry, University of California,
Berkeley, CA 94720 }

\begin{abstract}
Over the last two decades, scanning tunnelling microscopy (STM) has
become one of the most important ways to investigate the structure
of crystal surfaces. STM has helped achieve remarkable successes
in surface science such as finding the atomic structure of Si(111)
and Si(001). For high-index Si surfaces the information about the
local density of states obtained by scanning does not translate
directly into knowledge about the positions of atoms at the
surface. A commonly accepted strategy for identifying the atomic
structure is to propose several possible models and analyze their
corresponding {\em simulated} STM images for a match with the
experimental ones. However, the number of good candidates for the
lowest-energy structure is very large for high-index surfaces, and
heuristic approaches are not likely to cover all the relevant
structural models. In this article, we take the view that finding
the atomic structure of a surface is a problem of stochastic
optimization, and we address it as such. We design a general
technique for predicting the reconstruction of silicon surfaces
with arbitrary orientation, which is based on
parallel-tempering Monte Carlo simulations combined with an
exponential cooling. The advantages of the method are illustrated
using the Si(105) surface as example, with two main results: (a)
the correct single-step rebonded structure [e.g., Fujikawa {\em et
al.}, Phys. Rev. Lett. 88, 176101 (2002)] is obtained even when
starting from the paired-dimer model [Mo {\em et al.}, Phys. Rev.
Lett. 65, 1020 (1990)] that was assumed to be correct for many
years, and (b) we have found several double-step reconstructions
that have lower surface energies than any previously proposed
double-step models.
\end{abstract}
\maketitle

\section{Introduction}
Silicon surfaces are the most intensely studied crystal surfaces
since they constitute the foundation of the billion-dollar
semiconductor industry. Traditionally, the low-index surfaces such
as Si(001) are the widely used substrates for electronic device
fabrication. With the advent of nanotechnology, the stable
high-index surfaces of silicon have now become increasingly
important for the fabrication of quantum devices at length scales
where lithographic techniques are not applicable. Owing to their
grooved or faceted morphology, some high-index surfaces  can be
used as templates for the growth of self-assembled nanowires.
Understanding the self-organization of adatoms on these surfaces,
as well as their properties as substrates for thin film growth,
requires atomic-level knowledge of the surface structure. Whether
the surface unit cells are small [e.g., Si(113)] or large [such as
Si(5 5 12)], in general the atomic-scale models that were first
proposed were subsequently
contested:\cite{knal113,ranke113,dabrowski, 5512baski,
5512suzuki,5512takeguchi,mo,kds} the potential importance of
stable Si surfaces with certain high-index orientations sparked
many independent investigations, which led to different proposals
in terms of surface structure.

One of the most puzzling cases has been the (105) surface, which
appears on the side-facets of the pyramidal quantum dots obtained
in the strained layer epitaxy of Ge or Si$_{1-x}$Ge$_x$ ($x>0.2$)
on Si(001). Using STM imaging, Mo and coworkers proposed the first
model for this surface,\cite{mo} which was based on unrebonded
monatomic steps separated by small (two-dimer wide)
Si(001)-2$\times$1 terraces. Subsequently, Khor and Das Sarma
reported another possible (105) structure with a lower density of
dangling bonds.\cite{kds} However, the relative surface energy of
the two different reconstructions\cite{mo,kds} was not computed,
and the structure proposed in Ref.~\onlinecite{kds} had not, at
the time, replaced the widely accepted model\cite{mo} of Mo {\em
et al}. Only very recently it has been shown \cite{jap-prl,
jap-ss, italy-prl, apl, susc105} that the actual (105) structure
is made of  single-height rebonded steps (SR), which are strongly
stabilized by the compressive strains present in the Ge films
deposited on Si(001) \cite{italy-prl, apl} or Si(105).
\cite{jap-prl, jap-ss, susc105} Other high-index surfaces such as
Si(113) and Si(5 5 12) have sagas of their own,\cite{ranke113,
knal113, dabrowski,5512baski, 5512suzuki, 5512takeguchi} and only
in the former case there is now consensus\cite{dabrowski} about
the atomic structure.

The difficulty of finding the atomic structure of a surface is not
related to the resolution of the STM techniques, or to
understanding of the images obtained. After all, it is well-known
that STM gives information about the local density of states at
the surfaces and not necessarily about atomic coordinates.
\cite{luth} A common procedure for finding the reconstructions of
silicon surfaces consists in a combination of STM imaging and
electronic structure calculations as follows. Starting from the
bulk truncated surface and taking cues from the experimental data,
one proposes several atomic models for the surface
reconstructions. The proposed models are then relaxed using
density-functional or tight-binding methods, and STM images are
simulated in each case. At the end of the relaxation, the surface
energies of the structural models are also calculated. A match
with the experimental STM data is identified based on the relaxed
lowest-energy structures and their simulated STM images. This
procedure has long become standard and has been used for many
high-index orientations.\cite{knal113,dabrowski,jap-prl,
jap-ss,italy-prl,5512baski,si114} As described, the procedure is
heuristic, since one needs to rely heavily on physical intuition
when proposing good candidates for the lowest energy structures.
In the case of stable high-index Si surfaces, the number of
possible good candidates is rather large, and may not be exhausted
heuristically; thus, worst-case scenarios in which the most stable
models are not included in the set of "good candidates" are very
likely. On one hand, it has been recognized\cite{5512baski} that
the minimization of surface energy for semiconductor surfaces is
not controlled solely by the reduction of the dangling bond
density, but also by the amount of surface stress caused in the
process. On the other hand, intuitive reasoning can tackle (at
best) the problem of lowering the number of dangling bonds, but
cannot account for the increase in surface stress or for the
possible nanoscale faceting of certain
surfaces.\cite{baski001-111} For this reason, we adopt the view
that finding the structure of high-index Si surfaces is a problem
of stochastic optimization, in which the competition between the
saturation of surface bonds and the increase in surface stress is
intrinsically considered.

To our knowledge, a truly general and robust way of predicting the
atomic structure of semiconductor surfaces --understood as
finding the atomic configuration of a surface of any
arbitrary crystallographic orientation without experimental
input, has not been reported. It is not clear that such robust
atomic-scale predictions about semiconductor surfaces can even be ventured,
since theoretical efforts have been hampered by the lack of
empirical or semiempirical potentials that are {\em both fast and
transferable} for surface calculations. However, the long process
which lead to the discovery of the reconstruction of the (105)
surface\cite{mo,kds,jap-prl,jap-ss,italy-prl,apl,susc105}
indicates a clear need for a search methodology that does not rely
on human intuition. The goal of this article is to present a
strategy for finding the lowest-energy reconstructions for
an elemental crystal surface. While we hope that this strategy will
become a useful tool for many surface scientists, the extent of
its applicability remains to be explored. Our initial efforts will be focused
on the surfaces of silicon because of their utmost fundamental and
technological importance; nonetheless, the same strategy could be
applied for any other material surfaces provided suitable models
for atomic interactions are available.

\section{The Monte Carlo method}
\subsection{General considerations}
In choosing a methodology that can help predict the surface
reconstructions, we have taken into account the following
considerations. First, the number of atoms in the simulation slab
is large because it includes several subsurface layers in addition
to the surface ones. Moreover, the number of local minima of the
potential energy surface is also large, as it scales roughly
exponentially\cite{Sti83,Sti99} with the number of atoms involved
in the reconstruction; by itself, such scaling requires the use of
fast stochastic search methods. Secondly, the calculation of
interatomic forces is very expensive, so the method should be
based on Monte-Carlo algorithms rather than molecular dynamics.
Lastly, methods that are based on the modification of the
potential energy surface (PES) (such as the basin-hoping algorithm
\cite{basinhopping}), although very powerful in predicting global
minima, have been avoided as our future studies are aimed at
predicting not only the correct lowest-energy reconstructions, but
also the thermodynamics of the surface. These considerations
prompted us to choose the parallel-tempering Monte Carlo (PTMC)
algorithm \cite{Gey95,Huk96} for this study. Before describing in
detail the computational procedure and its advantages, we pause to
discuss the computational cell and the empirical potential used.
\begin{figure}
  \begin{center}
   \includegraphics[width=6cm]{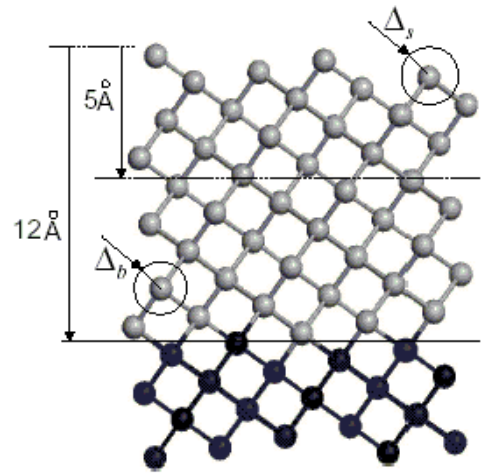}
  \end{center}
\caption{Schematic computational cell: the ``hot'' atoms (gray)
are allowed to move, while the bottom ones (black) are kept fixed
at their  bulk locations. Different maximum displacements
$\Delta_s$ and $\Delta_b$ are allowed for the atoms that are
closer to the surface and deeper in the bulk, respectively.}
\label{figgeom}
\end{figure}

The simulation cell has a single-face slab geometry with periodic
boundary conditions applied in the plane of the surface (denoted
$xy$), and no periodicity in the direction ($z$) normal to the
surface (refer to Fig.~\ref{figgeom}). The ``hot'' atoms from the
top part of the slab (corresponding to a thickness of 10--15~\AA)
are allowed to move, while the bottom layers of atoms are kept
fixed to simulate the underlying bulk crystal. Though highly
unlikely during the finite time of the simulation performed, the
evaporation of atoms is prevented by using a wall of infinite
energy that is parallel to the surface and situated 10~\AA \ above
it; an identical wall is placed at the level of the lowest fixed
atoms to prevent the (theoretically possible) diffusion of the hot
atoms through the bottom of the slab. The area of the simulation
cell in the $xy$-plane and the number of atoms in the cell are
kept fixed during each simulation; as we shall discuss in section
IV, these assumptions are not restrictive as long as we consider
all the relevant values of the number of atoms per area. Under
these conditions, the problem of finding the most stable
reconstruction reduces to the global minimization of the total
potential energy $V(\mathbf{x})$ of the atoms in the simulation
cell (here $\mathbf{x}$ denotes the set of atomic positions).  In
order to sort through the numerous local minima of the potential
$V(\mathbf{x})$, a stochastic search is necessary. The general
strategy of such search (as illustrated, for example, by the
simulated annealing technique\cite{Kir83, Kir84}) is to sample the
canonical Boltzmann distribution $ \exp\left[- V(\mathbf{x})/(k_B
T)\right] $ for decreasing values of the temperature $T$ and look
for the low-energy configurations that are generated.

In terms of atomic interactions, we are constrained to use
empirical potentials because the highly accurate ab-initio or
tight-binding methods are prohibitive. Since this work is aimed at
finding the {\em lowest} energy reconstructions for arbitrary
surfaces, the choice of the empirical potential is crucial, as
different interaction models can give different energetic ordering
of the possible reconstructions. Furthermore, the true structure
of the surface may not even be a local minimum of the potential
chosen to describe the interactions: it is the case, for example,
of the adatom-interstitial reconstructions\cite{dabrowski} of
Si(113), which are not local minima of the Stillinger-Weber
potential.\cite{stillweb} The work of Nurminen \emph{et
al.}\cite{landau-swt3-test} indicates that the most popular
empirical potentials for silicon\cite{stillweb,tersoff3} are not
suitable for finite-temperature simulations of surfaces. After
thorough numerical experimentation with several empirical
potentials, we have chosen to use the highly optimized empirical
potential (HOEP) recently developed by Lenosky \emph{et
al.}\cite{hoep} HOEP is fitted to a database of ab-initio
calculations that includes structural and energetic information
about small Si clusters, which leads to a superior transferability
to the different bonding environments present at the
surface.\cite{hoep}

\subsection{Advantages of the parallel tempering algorithm as a global optimizer}
The parallel tempering Monte Carlo method (also known as the
replica-exchange Monte-Carlo method) consists in running parallel
canonical simulations of many statistically independent replicas
of the system, each at a different temperature $T_1 < T_2 < \ldots
< T_N$. The set of $N$ temperatures $\{T_i,\ i=1,2,...N \}$ is
called a {\em temperature schedule}, or {\em schedule} for short.
The probability distributions of the individual replicas are
sampled with the Metropolis algorithm,\cite{Met53} although any
other ergodic strategy can be utilized. The key feature of the
parallel tempering method is that swaps between replicas of
neighboring temperatures $T_i$ and $T_j$ ($j = i \pm 1$) are
proposed and allowed with the conditional
probability\cite{Gey95,Huk96} given by
\begin{equation}
\label{eq:PTMCacc} \min\left\{1, e^{(1/T_j -
1/T_i)\left[V(\mathbf{x}_j)-V(\mathbf{x}_i)\right]/k_B}\right\},
\end{equation}
where $V(\mathbf{x}_i)$ represents the energy of the replica $i$
and $k_B$ is the Boltzmann constant. The conditional probability
(\ref{eq:PTMCacc}) ensures that the detailed balance condition is
satisfied and that the equilibrium distributions are the Boltzmann
ones for each temperature.

In the standard Metropolis  sampling\cite{Met53} of Boltzmann
distributions,  the probability that the Monte Carlo walker
escapes from a given local minimum decreases exponentially as the
temperature is lowered. In turn, the average number of Monte Carlo
steps needed for the walker to escape from the trapping local
minimum increases exponentially with the decrease of the
temperature, a scaling that makes the search for a global minimum
inefficient at low temperatures. To cope with this problem, the
parallel tempering algorithm takes advantage of the fact that the
Metropolis walkers running at higher temperatures have larger
probabilities of jumping over energy barriers. Parallel tempering
significantly decreases the time taken for the walker to escape
from local minima by providing an additional mechanism for jumping
between basins, namely the swapping of configurations between
replicas running at neighboring temperatures. Therefore, if a
given (low-temperature) replica  of the system is stuck in a local
minimum, the configuration swaps with walkers at higher
temperatures can provide that replica with states associated with
other basins (wells on the potential energy surface), ultimately
driving it into the global minimum.

Because of this swapping mechanism, parallel tempering enjoys
certain advantages (as a global optimizer) over the more popular
simulated annealing algorithm (SA).\cite{Kir83, Kir84} In order
for SA to be convergent (i.e. to reach the global optimum as the
temperature is lowered) the cooling schedule must be of the
form\cite{Gem84, Haj88}
\begin{equation}
\label{eq:SAlaw} T_i = \frac{T_{0}}{\log(i + i_0)}, \quad i \geq
1,
\end{equation}
where $T_{0}$ and $i_0$ are sufficiently large constants. Such a
logarithmic schedule is too slow for practical applications, and
faster schedules are routinely utilized.  Common SA cooling
schedules, such as the geometric or the linear ones,\cite{Kir83}
make SA non-convergent: the Monte Carlo walker has a non-zero
probability of getting trapped into minima other than the global
one.

The cooling schedule implied by Eq.~(\ref{eq:SAlaw}) is, of
course, asymptotically valid in the limit of low temperatures. In
the same limit, the PT algorithm allows for a geometric
temperature schedule.\cite{Sug00, Pre03} When the temperature
drops to zero, the system is well approximated by a
multidimensional harmonic oscillator and the acceptance
probability for swaps attempted between two replicas with
temperatures $T < T'$ is given by the incomplete beta function
law\cite{Pre03}
\begin{equation}
\label{eq:AcTT} Ac(T,T') \simeq \frac{2}{B(d/2,d/2)} \int_0^{1/(1
+ R)} \theta^{d/2 - 1}(1 - \theta)^{d/2 -1}d \theta \ ,
\end{equation}
where $d$ denotes the number of degrees of freedom of the system,
$B$ is the Euler beta function, and $R \equiv T' / T$. Since it
depends only on the temperature ratio $R$, the acceptance
probability (\ref{eq:AcTT}) has the same value for any arbitrary
replica running at a temperature $T_i$, provided that its
neighboring upper temperature $T_{i+1}$ is given by
$T_{i+1}=RT_{i}$. The value of $R$ is determined such that the
acceptance probability given by Eq.~(\ref{eq:AcTT}) attains a
prescribed value $p$, usually chosen greater that 0.5. Thus, the
(optimal) schedule that ensures a constant probability $p$  for
swaps between neighboring temperatures   is a geometric
progression:
\begin{equation}
 T_i = R^{i-1}T_{min},\quad 1 \leq i \leq N,
\label{eq:schedule}
\end{equation}
where $T_{min}=T_1$ is the minimum temperature of the schedule.
Though  more  research is required in order to better quantify the
relative efficiency of the two different algorithms SA and PT, it
is apparent from Eqs.~(\ref{eq:SAlaw}) and (\ref{eq:schedule})
that the parallel tempering algorithm is a global optimizer
superior to SA because it allows for a faster cooling schedule.
Direct numerical comparisons of the two methods have  confirmed
that parallel tempering is the superior optimization
technique.\cite{Mor03} The ideas of parallel tempering and
simulated annealing are not mutually exclusive, and in fact they
can be used together to design even more efficient stochastic
optimizers. As shown below, such a strategy that combines parallel
tempering and simulated annealing is employed for the present
simulations.
\begin{figure}
  \begin{center}
   \includegraphics[width=8cm]{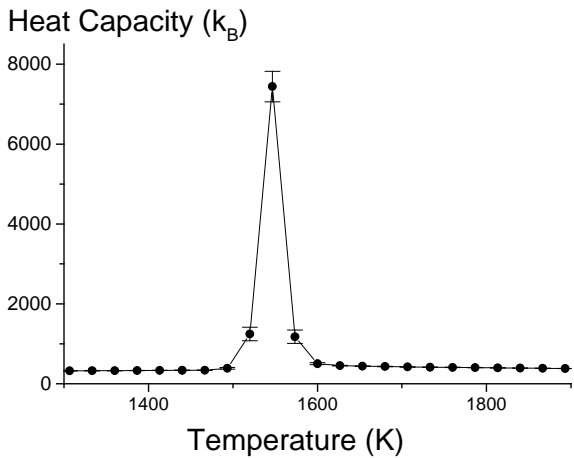}
  \end{center}
\caption{Heat capacity of a Si(105) slab plotted as a
function of temperature. The peak is located at 1550K;
in order to avoid recalculation of the heat
capacity for systems with different numbers of atoms and surface
orientations, we set $T_{max}=1600$K as the upper limit of the
temperatures range used in the PTMC simulations.}
\label{figheatcapacity}
\end{figure}
\subsection{Description of the algorithm}

The typical Monte Carlo simulation done in this work consists of
two main parts that are equal in terms of computational effort. In
the first stage of the computation, we perform a parallel
tempering run for a range of temperatures  $[T_{min},\ T_{max}]$.
The configurations of minimum energy are retained for each
replica, and used as starting configurations for the second part
of the simulation, in which each replica is cooled down
exponentially until the largest temperature drops below a
prescribed value.  As a key feature of the procedure, the parallel
tempering swaps are not turned off during the cooling stage. Thus,
we are using a combination of parallel tempering and simulated
annealing, rather than a simple cooling. Finally, the annealed
replicas are relaxed to the nearest  minima using a
conjugate-gradient algorithm. We now describe in detail the
stochastic minimization procedure. We shall focus, in turn, on
discussing the Monte Carlo moves, the choice of the temperature
range $[T_{min},\ T_{max}]$, and the total number of replicas $N$.

The moves of the hot atoms consist in small random displacements
with the $x,\ y,\ z$ components given by
\[
\Delta (2u_{\alpha}-1) \
\]
where $u_{\alpha}$ $(\alpha = x, y, z)$ are independent random
variables \cite{Mat98} uniformly distributed in the interval
$[0,1]$, and $\Delta$ is the maximum absolute value of the
displacement. We update the positions of the individual hot atoms
one at a time in a cyclic fashion.  Each attempted move is
accepted or rejected according to the Metropolis
logic.\cite{Met53} A complete cycle consisting in attempted moves
for all hot particles is called a \emph{pass} (or sweep) and
constitutes the basic computational unit in this work. We have
computed distinct acceptance probabilities for the hot atoms that
are closer to the surface (situated within a distance of
5~\AA\mbox{ } below the surface) and for the deeper atoms, the
movements of which are essentially small oscillations around the
equilibrium bulk positions. Consequently, as shown in
Fig.~\ref{figgeom}, we have employed two different maximal
displacements, $\Delta_s$ for the surface atoms, and $\Delta_b$
for the bulk-like atoms lying in the deeper subsurface layers. The
displacements $\Delta_s$ and $\Delta_b$ are tuned in the
equilibration phase of the simulation in such a way that the Monte
Carlo moves are accepted with a rate of 40\% to 60\%. This tuning
of the maximal displacements has been performed automatically by
dividing the equilibration phase into several blocks, computing
acceptance probabilities for each block, and increasing or
decreasing the size of the displacements $\Delta_{s,b}$ until the
acceptance probabilities  reached values between 40\% and 60\%.
The automatization is necessary because the optimal displacements
computed for replicas running at different temperatures have
different values. The maximal displacement $\Delta_s$ for the
surface atoms is found to be larger than the maximal displacement
for the bulk-like atoms. Though expected in view of the larger
mobility of the surface atoms, the difference between $\Delta_s$
and $\Delta_b$ is not substantial and the reader may safely employ
a single maximal displacement for all hot atoms at a given
temperature.

Parallel tempering configuration swaps are attempted between
replicas running at neighboring temperatures at every 10 passes in
an alternating manner, first with the closest lower temperature
then with the closest higher temperature. Exception make the two
replicas that run at end temperatures $T_1=T_{min}$ and
$T_N=T_{max}$, which are involved in swaps every 20 passes. The
range of temperatures $[T_{min}, T_{max}]$ and the temperature
schedule $T_1 < T_2 < \cdots < T_N $ have
been chosen as described below.

The maximum temperature $T_{max}$ must be high enough to ensure
that the corresponding random walker has good probability of
escaping from various local minima. However, as the temperature is
raised, increasingly more thermodynamic weight is placed on local
minima that have high energies compared to the global minimum.
Stillinger and Weber\cite{Sti83, Sti99} have argued that the
number of local minima increases exponentially with the
dimensionality of the system. As such, the probability that the
walker visits the basin of the global minimum significantly
decreases  with the increase of temperature. A very strong
decrease occurs at the melting point, beyond which most of the
configurations visited are associated with the liquid phase. The
basins of these configurations are unlikely to contain the global
minimum or, in fact, any of the low-energy local minima associated
with meaningful surface reconstructions. Therefore, the
high-temperature end must be set equal to the melting temperature.

The melting temperature of the surface slab can be determined from
a separate  parallel tempering simulation by identifying the peak
of the heat capacity plotted as a function of temperature. As
Fig.~\ref{figheatcapacity} shows, the melting temperature of a
Si(105) sample slab with $70$ hot atoms is about $1550~\text{K}$.
Rather than determining a melting temperature for each individual
system studied, we have employed a fixed value of $T_{max}=
1600~\text{K}$.  The melting temperature of the slab determined
here (Fig.~\ref{figheatcapacity}) is different from the value of
1250K reported for the bulk crystal:\cite{hoep} the discrepancy
is due to surface effects, finite-size effects, as well as to
the fact that the hot atoms are always in contact with the rigid
atoms from the bottom of the slab. Though we use $T_{max}=1600$K
for all simulations, we note that differences of 100K--200K in
the melting temperature of
the slab do not significantly affect the quality of the Monte
Carlo sampling. For most surfaces and system sizes of practical
importance, the value of $1600~\text{K}$ is in fact un upper bound
for the melting temperature; this may sometimes cause the one or
two walkers that run at the highest temperatures to be uncoupled
from the rest of the simulation, since they might sample amorphous
or liquid states. However, this loss in computational resources is
very small compared to the additional effort that would be
required by a separate determination of the heat capacity for each
surface slab used.
\begin{figure}
  \begin{center}
   \includegraphics[width=8.7cm]{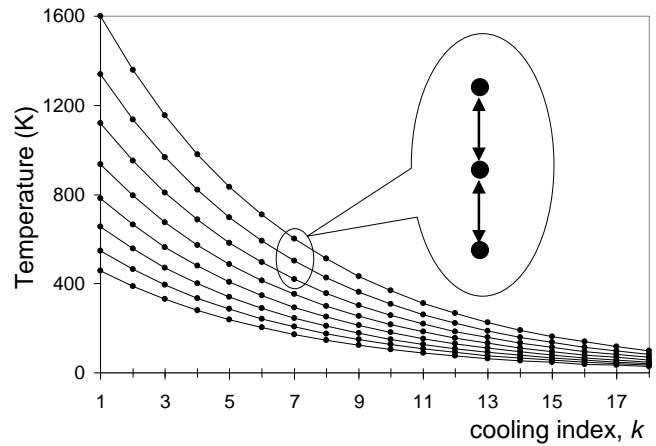}
  \end{center}
\caption{Exponential cooling of the $N=32$ Monte Carlo walkers
(replicas of the surface slab) used in the simulation. For
clarity, only eight walkers are shown (every fourth walker). The
cooling is performed in 18 steps: at each step the temperature is
modified by the same factor $\alpha=0.85$ for all walkers,
Eq.~(\ref{eq:cooling}). For every cooling step $k$, we have a
different parallel tempering schedule where each replica is
coupled to the walkers running at neighboring temperatures via
configuration swaps [Eq.~(\ref{eq:schedule}) with $R=4^{1/31}$].
This coupling is symbolized by the double-arrow lines in the
inset.} \label{figschedule}
\end{figure}

In theory, the lowest temperature $T_{min}$ should be set so low
that the walker associated with this temperature is virtually
localized in the basin associated with the global minima.
Nevertheless, obstacles concerning the efficient use of
computational resources prevent us from doing so. Numerical
experimentation has shown that a temperature of $T_{min} =
400~\text{K}$ is low enough that only local minima associated with
realistic surface reconstructions are frequently visited. A
further selection among these local minima is performed in the
second part of the Monte Carlo simulation, when all temperatures
of the initial schedule $\{T_i, \ i=1,2,...N \}$ are gradually
lowered to values below $100~\text{K}$; as it turns out,  this
combination of parallel tempering and simulated  annealing makes
optimal use of computational resources. Below the melting point
the heat capacity  of the surface slab is almost constant and well
approximated by the capacity of a multidimensional harmonic
oscillator (refer to Fig.~\ref{figheatcapacity}). In these
conditions, the acceptance probability for swaps between
neighboring temperatures $T$ and $T'$ is given by
Eq.~(\ref{eq:AcTT}) (see also Ref. \onlinecite{Pre03}). It follows
that the optimal temperature schedule on the interval $[T_{min},
T_{max}]$  is the geometric progression (\ref{eq:schedule}), where
\[
R = (T_{max}/T_{min})^{1/[N(d,p)-1]}.
\]
We have written $N\equiv N(d,p)$ to denote the smallest number of
replicas that guarantees a swap acceptance probability of at least
$p$ for a system with $d$ degrees of freedom. Since the best way
to run PTMC  calculations  is to use one processor for each
replica  of the system, the feasibility of our simulations hinges
on  values of $N(d,p)$ that translate directly into available
processors. The number of walkers $N(d,p)$ can be
estimated\cite{Pre03} by
\begin{equation}
\label{eq:minimumN} N(d,p) =  \left[d^{1/2}
\frac{\sqrt{2}\ln(T_{max}/T_{min})}{4\text{erf}^{-1}(1-p)}\right]
+ 2,
\end{equation}
where $[x]$ denotes the largest integer smaller than $x$, and
erf$^{-1}$ is the inverse error function. Based on
Eq.~(\ref{eq:minimumN}), we have used $N=32$ walkers for all
simulations, which ensures a swap acceptance ratio greater than
$p=0.5$ for any system with less than 300 hot atoms, $d<900$. The
first part of all Monte Carlo  simulations performed in the
present article consists of a number of $36\times 10^{4}$ passes
for each replica, preceded by $9\times 10^4$ passes allowed for
the equilibration phase. When we retained the configurations of
minimum energy, the equilibration passes have been discarded so
that any memory of the starting configuration is lost.

We now describe the second part of the Monte Carlo simulation,
which consists of a combination of simulated annealing and
parallel tempering. At the $k$-th cooling step, each temperature
from the initial temperature schedule $\{ T_i, i=1,2,..N \}$ is
decreased by a factor which is independent of the index $i$ of the
replica, $T_{i}^{(k)} = \alpha_{k} T_{i}^{(k-1)}.$ Because the
parallel tempering swaps are not turned off, we require that at
any cooling step $k$ all $N$ temperatures must be modified by the
same factor $\alpha_{k}$ in order to preserve the original swap
acceptance probabilities. The specific way in which $\alpha_k$
depends on the cooling step index $k$ is determined by the kind of
annealing being sought. In this work we have used a cooling
schedule of the form
\begin{equation}
\label{eq:cooling} T_{i}^{(k)} = \alpha T_{i}^{(k-1)}=\alpha
^{k-1}T_i \ \ \ \ (k \geq 1),
\end{equation}
where $T_i\equiv T_i^{(1)}$ and $\alpha$ is determined such that
the temperature intervals $[T_{1}^{(k-1)}, T_{N}^{(k-1)}]$ and
$[T_{1}^{(k)},T_{N}^{(k)}]$ spanned by the parallel tempering
schedules before and after the $k$-th cooling step overlap by
$80\%$. This yields a value for $\alpha$ given by $(0.2 T_{min}+
0.8 T_{max})/{T_{max}} = 0.85. $ We have also tested values of
$\alpha$ larger than 0.85, and did not find any significant
improvements in the quality of the sampling.

The reader may argue that the use of an exponential annealing
[Eq.~(\ref{eq:cooling})] is not the best option for attaining the
global energy minimum of the system. Apart from the theoretical
considerations discussed in the preceding subsection that only a
logarithmic cooling schedule would ensure convergence to the
ground state,\cite{Gem84,Haj88}it is known that the best annealing
schedules for a given computational effort oftentimes involve
several cooling-heating cycles. We emphasize that in the present
simulations, the most difficult part of the sampling is taken care
of by the initial PTMC run. In addition,  since the configuration
swaps are not turned off during cooling (refer to
Fig.~\ref{figschedule}), the Monte-Carlo walkers are subjected to
cooling-heating cycles through the parallel tempering algorithm.

The purpose of the annealing (second part of the simulation) is to
cool down the best configurations determined by the initial
parallel tempering in a way that is more robust than the mere
relaxation into the nearest local minimum. If the initial PTMC run
is responsible for placing the system in the correct funnels
(groups of local minima separated by very large energy barriers),
the annealing part of the simulation takes care of jumps between
local minima separated by small barriers within a certain funnel.
For this reason, the annealing is started from the configurations
of minimum energy determined during the first part. The cooling is
stopped when the largest temperature in the parallel tempering
schedule drops below $100$K. This criterion yields a total of 18
cooling steps, with $2\times 10^4$ MC passes per replica performed
at every such step.

Each cooling step is preceded by $5\times 10^3$ equilibration
passes, which are also used for the calculation of new maximal
displacements $\Delta_s$ and $\Delta_b$, as these displacements
depend on temperature and must be recomputed. In fact, each
cooling step is a small-scale version of the first part of the
simulation. The only difference is that the cooling steps are
\emph{not} started from the configurations of minimum energy
determined at the preceding cooling steps. (Otherwise, because the
number of passes for a given step is quite small, the walkers
might not have time to escape from some spurious local minima and
we would end up restarting them over again from the respective
minima.)

The third and final part of the minimization procedure is a
conjugate-gradient optimization of the last configurations
attained by each replica. The relaxation is necessary because we
aim to classify the reconstructions in a way that does not depend
on temperature, so we compute the surface energy at zero Kelvin
for the relaxed slabs $i,\ i=1,2,...N$. The surface energy
$\gamma$ is defined as the excess energy (with respect to the
ideal bulk configuration) introduced by the presence of the
surface:
\begin{equation}
\gamma =\frac{E_m-n_m e_b}{A}
\end{equation}
where $E_m$ is the potential energy of the $n_m$ atoms that are
allowed to move, $e_b=4.6124$eV is the bulk cohesion energy given
by HOEP, and $A$ is the surface area of the slab.

\section{Results for the Si(105) surface}
We have tested the method for a variety of surface orientations,
such as (113), (105) and (5 5 12). In this section we are
presenting results for Si(105), a choice that was determined by
the ubiquity of the (105) orientation on the side facets of the
pyramidal quantum dots obtained in the heteroepitaxial deposition
of Ge and Si-Ge alloys on Si(001). Recent experimental and
theoretical work on the atomic structure of (105) surfaces
\cite{jap-prl,jap-ss,italy-prl,apl,susc105} provides a strong
testing ground for the current investigations. In order to assess
the versatility of the method and to provide a direct comparison
with a previous heuristic study \cite{susc105} of the (105)
reconstructions, we start our PTMC simulations from each of the
structures found in Ref.~\onlinecite{susc105}.
\begin{figure*}
  \begin{center}
   \includegraphics[width=15.00cm]{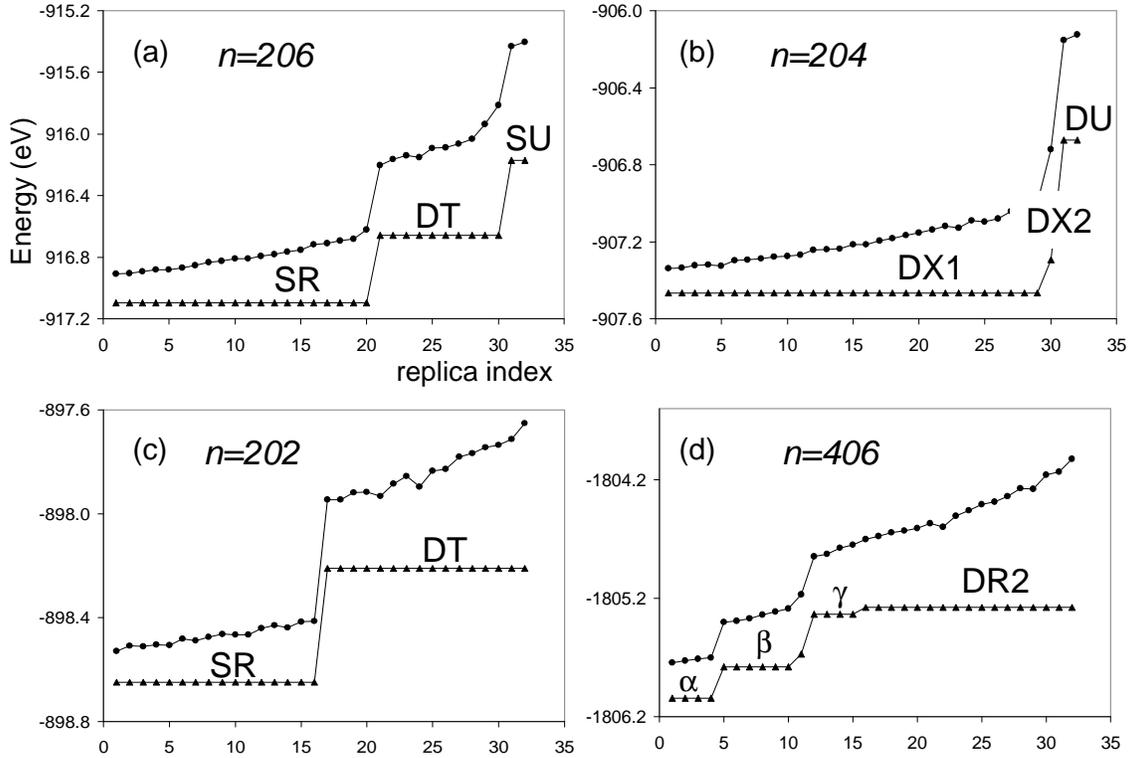}
  \end{center}
\caption{Total energies of the 32 replicas of the Si(105)
computational slab at the end of the cooling sequence (circles),
and after the subsequent conjugate-gradient relaxation
(triangles). The PTMC procedure has been started with all the
replicas in the same configuration taken from the set reported in
Ref.~\onlinecite{susc105}: SU(a), DU(b), DR(c), DR2(d).
The lowest-energy configurations depend on the total number of atoms
$n$, which is indicated in each panel. Six new
double-step structural models are found, denoted by DT, DX1, DX2,
DR2$\alpha$, DR2$\beta$ and DR2$\gamma$, with surface energies
smaller than those of the corresponding starting structures. }
\label{figharvest}
\end{figure*}
\begin{figure*}
  \begin{center}
   \includegraphics[width=13cm]{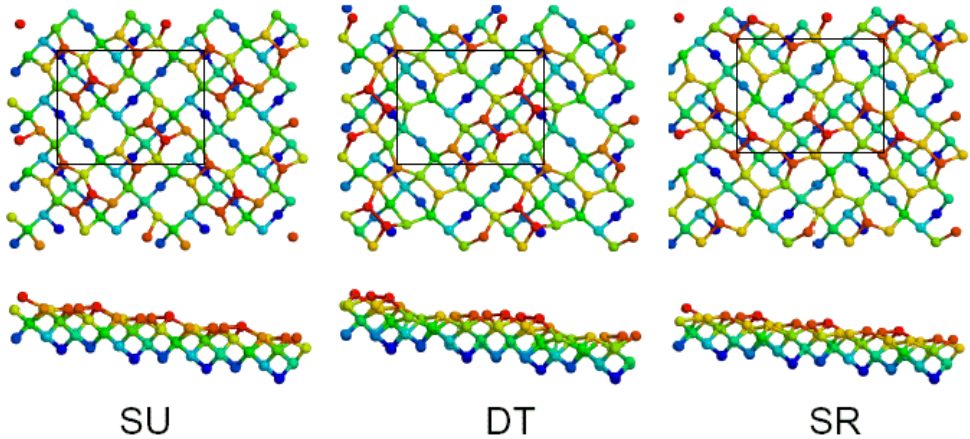}
  \end{center}
\caption{Si(105) reconstructions obtained when starting from the
SU model: SU, DT and SR. The DT structure is a novel
double-stepped structure retrieved by replicas running at
intermediate temperatures (see also Fig.~\ref{figharvest}(a)). The
single-step rebonded structure \cite{kds,
jap-prl,jap-ss,italy-prl,apl,susc105} (SR) is the global optimum.
The rectangle represents the surface unit cell, which is the same as the
as the periodic supercell used in the simulations. Atoms are
rainbow-colored according to their coordinate along the [105]
direction, with the red atoms being at the highest positions.} \label{figSUDTSR}
\end{figure*}
To establish the nomenclature for the discussion to follow, we
recall that the structures were labelled  by SU, SR, DU, DU1, DR,
DR1, and DR2, where the first letter denotes the height of the
steps (single S, or double D), the second letter indicates whether
the step is rebonded (R) or unrebonded (U) and the digit
distinguishes between different structures that have the same
broad topological features.\cite{susc105} These reconstructions
have different numbers of atoms and different linear dimensions of
the periodic cell. The dimensions of the cell are chosen $2a\times
a\sqrt{6.5}$ ($a=5.431$\AA \ is the bulk lattice constant of Si)
for all the models considered except DR2, whose topology requires
a periodic cell of $2a\times 2a\sqrt{6.5}$. The thickness of the
slab corresponds to two unit cells in the $z$ direction, with a
maximum of 208 atoms, of which only about half are allowed to
move.

The results of the PTMC simulations for the Si(105) surface are
plotted in Fig.~\ref{figharvest}, which shows the total energy for
each of the $N=32$ replicas at the end of the cooling procedure
(circles) and after the conjugate-gradient relaxation (triangles).
Figs.~\ref{figharvest}(a), (b), (c) and (d) show the total
energies of the reconstructions obtained starting from the SU, DU,
DR and DR2 models, respectively. In each case, we have obtained at
least two structures with lower surface energies than the starting
configurations, which we discuss in turn.
\begin{figure*}
  \begin{center}
   \includegraphics[width=15.00cm]{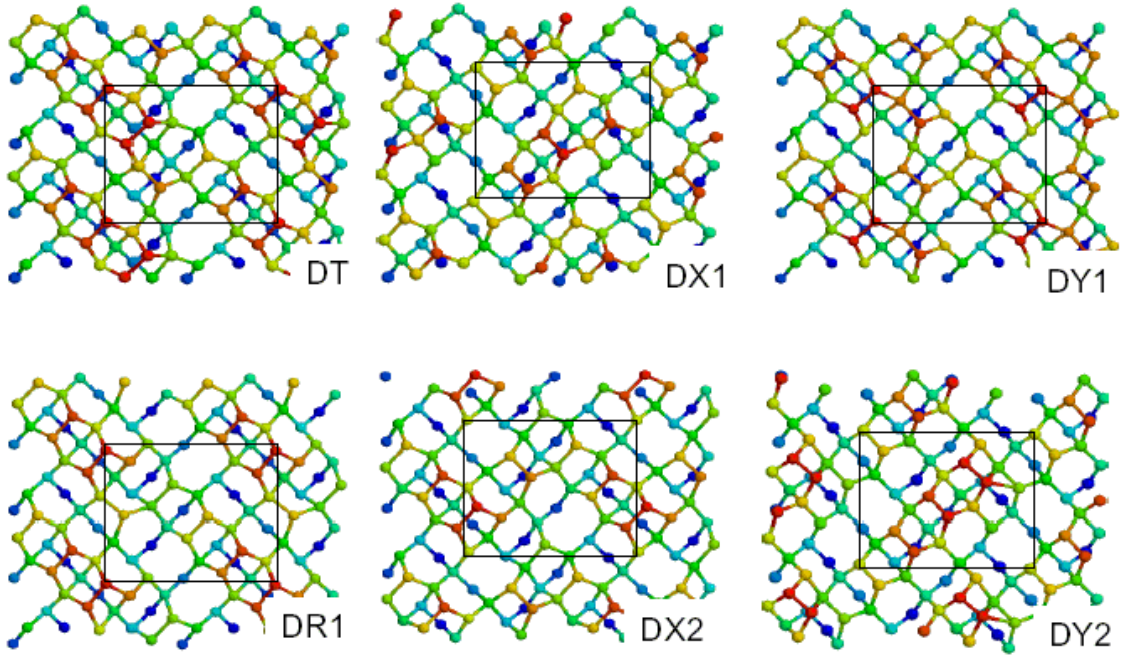}
  \end{center}
\caption{Double-step reconstructions of Si(105) with periodic
cells (rectangles shown) of dimensions $2a\times a\sqrt{6.5}$. The
color scheme is the same as in Fig.~\ref{figSUDTSR}. Except for
DR1, all other structures are new.} \label{figDTRX}
\end{figure*}

Fig.~\ref{figharvest}(a) shows that the (starting) SU
structure\cite{mo} is found only by the two replicas running at
the highest temperatures, while colder walkers find a novel
double-stepped structure, termed here ``transitional'' (DT). At
even lower temperatures, the double-steps of the DT reconstruction
unbunch into single-height rebonded (SR) steps; the three
different configurations that correspond to the energies plotted
in Fig.~\ref{figharvest}(a) are shown in Fig.~\ref{figSUDTSR}.
Therefore, the correct SR structure
\cite{jap-prl,jap-ss,italy-prl,apl,susc105} is retrieved even when
starting from the topologically different SU model. The usefulness
of this PTMC simulation becomes apparent if we recall that the SU
structure was widely believed to be correct for more than a decade
after its publication. As we shall see, the ground state obtained
in our stochastic search is independent of the initial
configuration. The only condition for finding the reconstruction
with the lowest surface energy is to prescribe the correct number
of atoms and the correct dimensions for the simulation slab. We
will address these practical aspects in the next section; for now,
we continue to describe the results obtained for different numbers
of atoms in the computational slab.

The simulation that starts from the DU model finds two distinct
rebonded structures, denoted by DX1 and DX2 in
Fig.~\ref{figharvest}(b). Both these structures are characterized
by the presence of single dimers at the location of steps (see
Fig.~\ref{figDTRX}), which reduces the number of dangling bonds
per unit area from 6$db/a^2\sqrt{6.5}$ (starting structure DU) to
5$db/a^2\sqrt{6.5}$. The DX1 reconstruction is the most stable,
and it is obtained in all but three replicas of the system.

\begin{figure}
  \begin{center}
   \includegraphics[width=8.70cm]{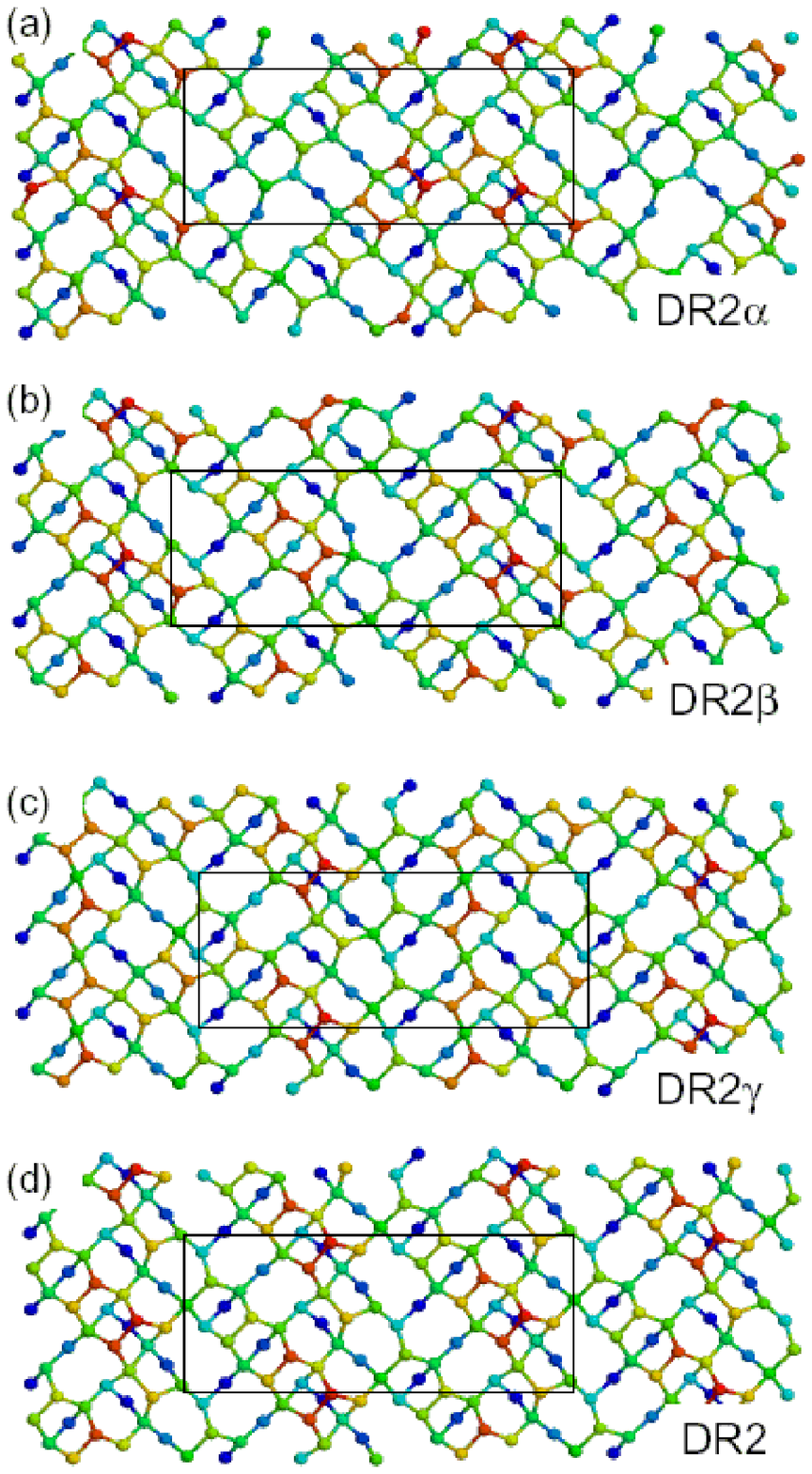}
  \end{center}
\caption{Double-step reconstructions of Si(105) with periodic
cells (rectangles shown) of $2a\times 2a\sqrt{6.5}$. Although the
starting structure [the DR2 model\cite{susc105} shown in (d)] has
a reasonably low dangling bond density ($5 db/a^2\sqrt{6.5}$), the
Monte Carlo simulation has retrieved three more reconstructions,
all having smaller surface energies (refer to
Table~\ref{table_gamma_smcell}). These novel structures [shown in
figs. (a)--(c)] are labelled by DR2$\alpha$, DR2$\beta$,
DR2$\gamma$. The atoms are rainbow-colored as indicated in
Fig.~\ref{figSUDTSR}.} \label{figDR2}
\end{figure}

Although it has a small density of dangling bonds, the DR
structure has large surface energy due to the $\sqrt{2} \times 1$
terrace reconstruction.\cite{susc105} Since the density of
dangling bonds is the lowest possible (4$db/a^2\sqrt{6.5}$), the
minimization of surface energy is dictated by the reduction of
surface stress. Unlike the case of SU and DU structures (described
above), not a single replica have retained the starting model DR.
Instead, the DT and SR structures are retrieved (refer to
Fig.~\ref{figharvest}(c)). When starting from the DR2 structure we
obtain at least three low energy structures denoted by
DR2$\alpha$, DR2$\beta$ and DR2$\gamma$ (Fig.~\ref{figDR2}), which
have not been previously proposed in Refs. \onlinecite{susc105},
\onlinecite{china-Si105}, or elsewhere. Owing to a larger area of
the slab, portions of the newly reconstructed unit cells have
patches that resemble the models obtained in prior simulations. In
particular, the atomic scale features of the steps on DR2$\alpha$
are very similar to those of the SR structure, a similarity that
reflects in the very small relative surface energy of the two
models ($\approx 1.6$meV/\AA$^2$\ ).

We note that the simulations described have a total number of
atoms that is between $n=202$  and $n=206$
(Fig.~\ref{figharvest}(a) and (c)) per $2a^2\sqrt{6.5}$ area. To
cover all the possibilities for intermediate numbers of atoms, we
also perform a simulation with $n=205$; this value of $n$ does not
correspond to any of the models reported in Ref.
\onlinecite{susc105}, and the parallel tempering run is started
from a bulk-truncated configuration. In this case two new
structures are found; these structures are named DY1 and DY2 and
shown in Fig.~\ref{figDTRX}. [The letters X and Y appearing in
DX1, DX2, DY1, DY2 (all denoting double-stepped rebonded
structures, Fig.~\ref{figDTRX}) do not stand for particular words,
they are simply intended to unambiguously label the structures in
a way that does not complicate the notation.] While for the DY1
model the rebonding is realized via bridging bonds,\cite{susc105}
in the case of DY2 we find unexpected topological features such as
fully saturated surface atoms and over-coordinated bulk atoms.
Even though these structural units (seen in the DY2 panel of
Fig.~\ref{figDTRX}) reduce the number of dangling bonds, they also
create high atomic-level stresses which make the DY2
reconstruction relatively unfavorable.

We have also performed PTMC simulations with SR, DR1 and DU1 as
initial configurations, but have not obtained any other novel
reconstructions. We found that SR and DR1 are the global energy
minima corresponding to 206 atoms and 203 atoms, respectively. The
DU1 structure\cite{china-Si105} (202 atoms) has lead to the same
reconstructions as the SU model (206 atoms). This result indicates
a periodic behavior of the surface energy as a function of the
total number of atoms, which will be discussed next.

\begin{figure}
  \begin{center}
   \includegraphics[width=7cm]{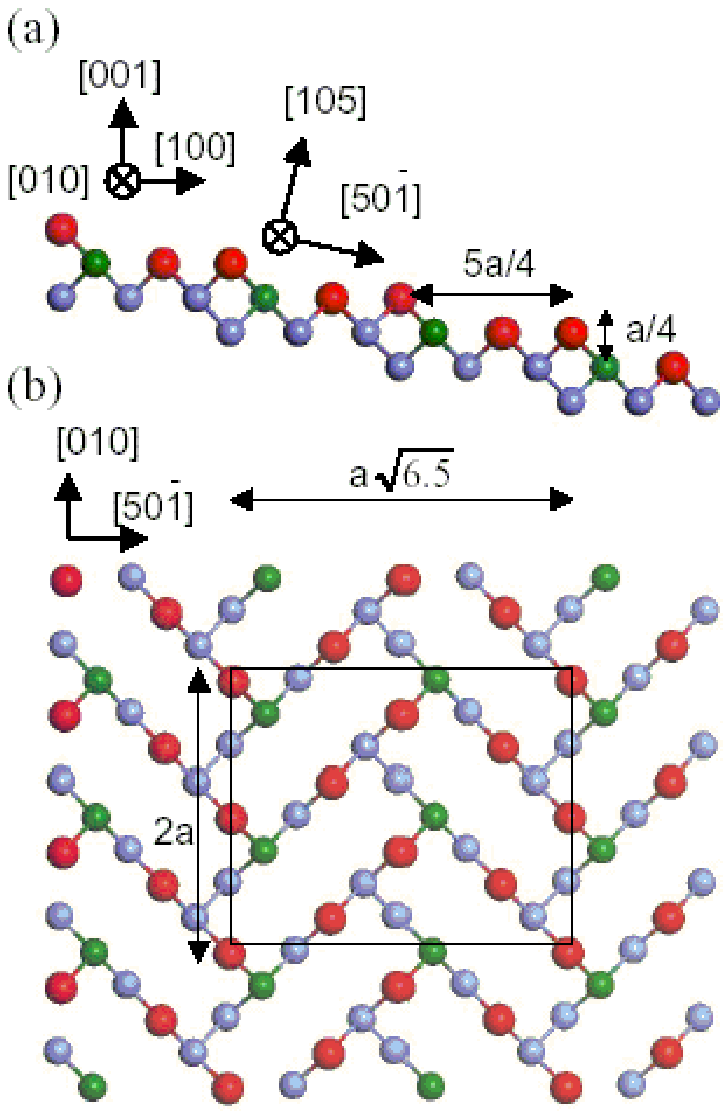}
  \end{center}
\caption{Atomic structure of the bulk truncated Si(105) surface,
viewed from the side (a) and from the top (b). The rectangle of
dimensions $2a\times a\sqrt{6.5}$ marks the periodic cell used in
most of the simulations, and contains two unit cells of the
bulk-truncated surface. For clarity, only a single subsurface
(001) layer is shown. In this picture (unlike in
Figs.~\ref{figSUDTSR}, \ref{figDTRX} and \ref{figDR2}) atoms are
colored according the their number of dangling bonds ($db$) before
reconstruction: red$ = 2db$, green$ = 1db$ and blue$ = 0db$.}
\label{figbulktruncated}
\end{figure}

\section{Discussion}
\begin{figure}
  \begin{center}
   \includegraphics[width=8.7cm]{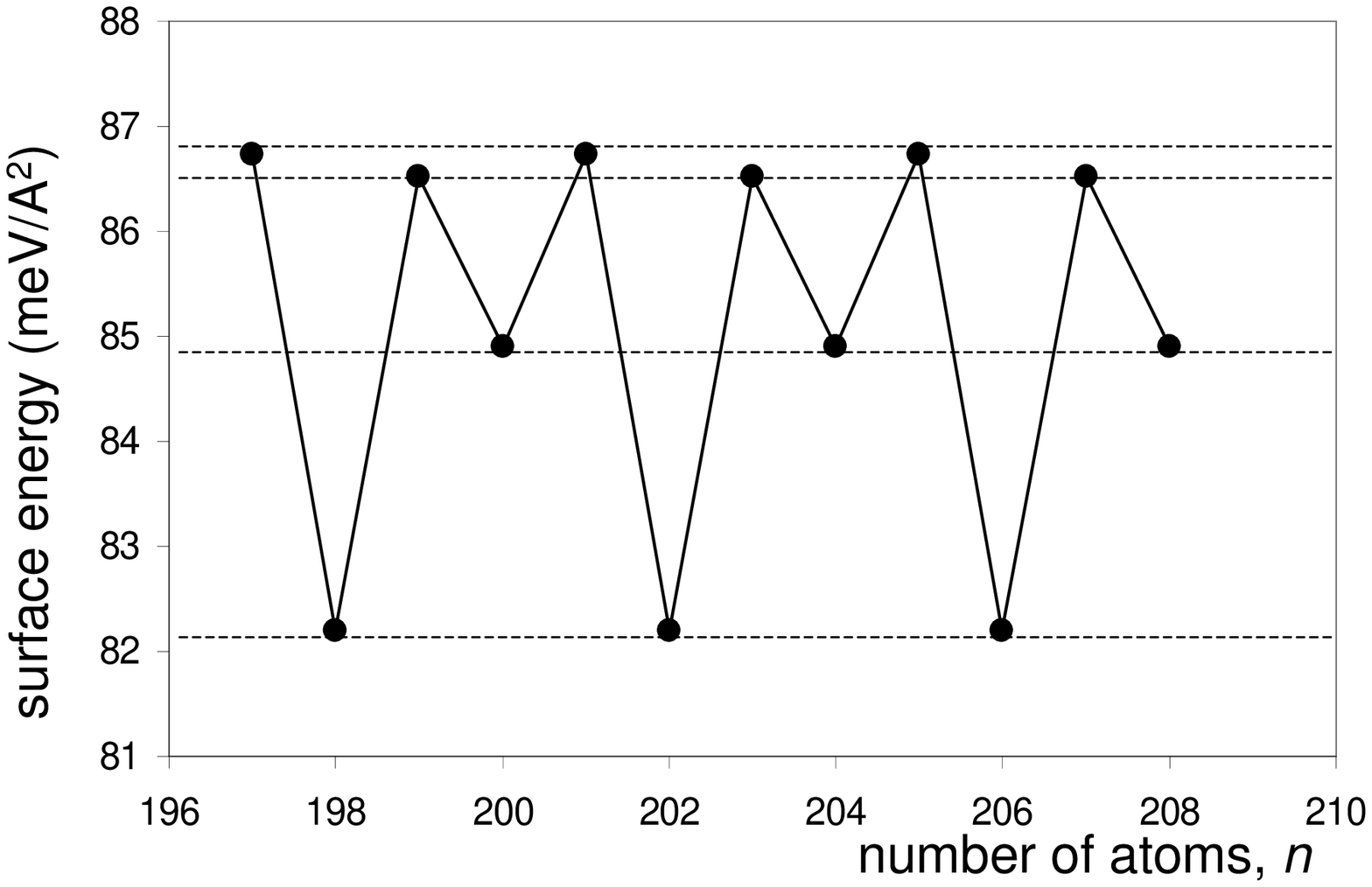}
  \end{center}
\caption{Surface energy of the global minimum structure plotted
versus the total number of atoms $n$ in the simulation slab. Even
though there are twelve under-coordinated atoms in each
bulk-truncated periodic cell (refer to
Fig.~\ref{figbulktruncated}), the values of the surface energy
repeat at intervals of $\Delta n =4$. The underlying bulk
structure reduces the number of distinct global minima to four.}
\label{figperiodicsurfene}
\end{figure}

%
%
To further test that the lowest energy states for given number of
atoms are independent of the initial configurations, we have
repeated all the calculations using bulk-truncated surface slabs
(Fig.~\ref{figbulktruncated}) instead of reconstructed ones. We
have varied the number of atoms $n$ in the simulation cell between
196 and 208, where the latter corresponds to four bulk unit cells
of dimensions $a\sqrt{6.5}\times a \times a\sqrt{6.5}$ stacked two
by two in the [010] and [105] directions. For the cases with
$n<208$, we have started the PTMC simulations from structures
obtained by taking out a prescribed number atoms from random
surface sites, and have found the same ground state irrespective
of the locations of the removed atoms. For values of $n$ equal to
202, 203, 204 and 206, the ground states (global minima) are also
the same as the ones obtained from the reconstructed models DR,
DR1, DU, and SU, respectively. Furthermore, we have tested that
even when removing arbitrary {\em subsurface} atoms the simulation
retrieves the same ground states without increasing the
computational effort. This finding speaks for the quality of the
Monte Carlo  sampling and gives confidence in the predictive
capabilities of the method described in section II. The lowest
surface energies obtained at the end of the numerical procedure
are shown in Fig.~\ref{figperiodicsurfene} as a function of the
number of atoms in the simulation cell.
As illustrated in Fig.~\ref{figperiodicsurfene}, the simulation
finds the same ground states at periodic intervals of $\Delta n =
4$. At first sight, this is somewhat surprising given that the
number of under-coordinated surface atoms in a bulk-truncated cell
of dimensions $2a \times a\sqrt{6.5}$ is twelve (refer to
Fig.~\ref{figbulktruncated}). The reduced periodicity of the
surface energy with the number of atoms in the supercell is due to
the underlying crystal structure, which lowers the number of
symmetry-distinct global minima to only four. Thus, we have
considered all possibilities in terms of numbers of atoms in a
simulation slab of area $2a^2\sqrt{6.5}$. The surface energies of
the optimal reconstructions for relevant values of $n$, as well as
those of some higher-energy structures, are collected in
Table~\ref{table_gamma_smcell}. As shown in the table, the global
minimum of the surface energy of Si(105) is obtained for the
single-height rebonded structure SR. While this finding is in
agreement with recent reports,\cite{jap-prl, jap-ss, italy-prl,
apl, susc105} it is the result of an exhaustive search rather than
a comparison between two\cite{jap-prl, jap-ss,italy-prl,apl} or
more\cite{susc105} heuristically proposed structures.
\begin{table}
\begin{center}
\begin{tabular}{r l c c c}
\hline \hline
$n$   &  Structure   &  Bond counting        & HOEP  &   TB  \\
\     &  \           &  ($db/a^2\sqrt{6.5}$) & (meV/\AA$^2$) & (meV/\AA$^2$)   \\
\hline

206  & SR   & 4 & 82.20  & 82.78 \\
\    & DT   & 4 & 85.12  &  \    \\
\    & SU   & 6 & 88.35  & 83.54 \\

205  &  DY1 & 5 & 86.73 & \ \\
\    &  DY2 & 4.5 & 88.61 & \ \\

204  &  DX1 & 5 & 84.90 & \ \\
\    &  DX2 & 5 & 86.04 & \ \\
\    &  DU  & 6 & 90.18 & 84.84 \\

203  &  DR1 & 5 & 86.52 & 85.22 \\

2$\times$203  &  DR2$\alpha$ & 4.5 & 83.77 & \  \\
2$\times$203  &  DR2$\beta$  & 4.5 & 84.64 & \  \\
2$\times$203  &  DR2$\gamma$ & 4.5 & 86.15 & \  \\
2$\times$203  &  DR2         & 5 & 86.34 & 83.48 \\

\hline \hline
\end{tabular}
\caption{Surface energies of different Si(105) reconstructions,
calculated using the HOEP interatomic potential.\cite{hoep} The
structures are grouped according to the number of atoms $n$ in the
simulation cell. Atomic configurations of selected reconstructions
are shown in Figs.~\ref{figSUDTSR}, \ref{figDTRX} and
\ref{figDR2}. The third column shows the number of dangling bonds
($db$) per unit area, expressed in units of $a^2 \sqrt{6.5}$. The
last column indicates the tight-binding\cite{tbmd}(TB) values
reported in Ref.~\onlinecite{susc105}.} \label{table_gamma_smcell}
\end{center}
\end{table}

%
%
From Table~\ref{table_gamma_smcell} we also note that the SR and
the DR2$\alpha$ structures have surface energies that are within
1.6meV/\AA$^2$ from one another. This gap in the surface energy
of the two models (SR and DR2) is smaller than the expected
accuracy of relative surface energies determined by an
empirical potential. Therefore, it is very likely that these
two reconstructions can both be present on the same surface under
laboratory conditions. As recently pointed out,\cite{susc105} the
coexistence of several configurations with different topological
features but similar surface energies gives rise to the atomically
rough and disordered aspect\cite{tomitori, china-Si105} of the
Si(105) surface. The surface energies computed using HOEP for
various rebonded structures (Table~\ref{table_gamma_smcell}) are
close to the values obtained previously\cite{susc105} at the
tight-binding level.\cite{tbmd} For the unrebonded structures (SU
and DU), the differences between the HOEP values and the
tight-binding ones are larger: this discrepancy is caused by the
inability of the HOEP interaction model to capture the tilting of
the surface dimers, which is an important mechanism for the
relaxation of these unrebonded configurations. Despite this
shortcoming, we have found that the HOEP potential is accurate
enough to predict the correct bonding topology of the global
minimum reconstructions for a variety of surface orientations. If
a comparison with experimental STM images is desired, further
geometry optimizations are necessary at the level of electronic
structure methods: these calculations would have to consider
different tiltings of the surface bonds, and in each case the
simulated image is to be compared with the experimental one. Thus,
even for surfaces where dimer tilting is important, the Monte
Carlo simulation based on the HOEP interaction model\cite{hoep}
can still serve as a very efficient tool to find good candidates
for the lowest energy structures.

Two practical issues have to be addressed when using PTMC
simulations for surface structure prediction. First consideration
is related to the size of the computational cell. If a periodic
surface pattern exists, the lengths and directions of the surface
unit vectors can be determined accurately through experimental
means (e.g., STM). In those cases, the periodic lengths of the
simulation slab should simply be chosen the same as the ones found
in experiments. On the other hand, when the surface does not have
two-dimensional periodicity (as it is the case of unstrained
Si(105) surface\cite{tomitori, china-Si105}), or when experimental
data is not available, one should systematically test
computational cells with periodic vectors that are low-integer
multiples of the unit vectors of the bulk truncated surface; the
latter unit vectors can be easily computed from the knowledge of
crystal structure and surface orientation. Secondly, the number of
atoms in the simulation cell is not {\em a priori} known, and
there is no simple criterion to find the set of numbers that yield
the lowest surface energy for a slab with arbitrary orientation.
Adapting the algorithm presented in section II for a
grand-canonical ensemble is somewhat cumbersome, as one would have
to consider efficiently the combination of two different types of
Monte Carlo moves: the small random displacements of the atoms
(continuous) and the discrete processes of adding or removing
atoms from the simulation slab. The problem of finding the correct
number of atoms in the computational cell is not new, as it also
appears, for example, in classic algorithms for predicting the
bulk crystal structure.\cite{jap-pr} As shown above for the case
of Si(105), a successful way to deal with this problem is to
simply repeat the simulation for systems with consecutive numbers
of atoms, and look for a periodic behavior of the surface energy
of coldest replicas as a function of the number of particles in
the computational cell. Note that if the thickness of the slab is
sufficiently large, such periodicity of the lowest surface energy
with respect to the number of atoms in the supercell is {\em
guaranteed} to exist: in the worst case, the periodicity will
appear when an entire atomic layer has been removed from the
simulation cell.

\section{Concluding Remarks}
%
In conclusion, we have developed and tested a stochastic method
for predicting the atomic configuration of silicon surfaces. If
suitable empirical models for atomic interactions are available,
this method can be straightforwardly applied for the determination
of the structure of any crystallographic surface of any other
material. Using the example of Si(105), we have shown that the
PTMC search is superior to heuristic approaches because it ensures
that the topology corresponding to the lowest surface energy is
considered in the set of good possible structural models. We have
performed an exhaustive search for different numbers of atoms in
the simulation cell and have found that the global minimum of the
(105) surface energy is the single-height rebonded model SR, in
agreement with recent studies.\cite{jap-prl, jap-ss, italy-prl,
apl, susc105} The experiments of Zhao {\em et
al.}\cite{china-Si105} indicated that double-stepped structures
are present on the unstrained Si(105) surface: our simulations
indeed have found double-stepped models with surface energies that
are close to the surface energy of the optimal SR reconstruction.
In addition, these double-stepped models (termed DR2$\alpha$,
DR2$\beta$, and DR2$\gamma$) are energetically more favorable than
the double-stepped structures proposed in Refs.
\onlinecite{china-Si105} and \onlinecite{susc105}.

We would like to comment on the key role played by the empirical
potential used in the present simulations. A highly transferable
interatomic potential is required for a satisfactory energetic
ordering of different reconstructions. While we would not expect
any empirical potential to accurately reproduce the relative
surface energies of all the reconstructions found, we can at least
expect that the chosen potential correctly predicts the bonding
topology for well-known surface reconstructions. In this respect,
the HOEP model\cite{hoep} proved superior to the most widely used
interatomic potentials.\cite{stillweb, tersoff3} Given this
comparison, the results presented here would represent a
validation of the work\cite{hoep} towards more transferable
potentials for silicon. We also hope that these results would
stimulate further developments of interatomic potentials for other
semiconductors.\cite{feigao}

With the exception of Si(105), Si(113)\cite{dabrowski} and
(likely) Si(114),\cite{si114} the atomic structure of other stable
high-index silicon surfaces has not been elucidated, although a
substantial body of STM images has accumulated to
date.\cite{si-highindex} A similar situation exists for Ge
surfaces as well.\cite{ge-highindex} The methodology presented in
this article can be used (either directly or in combination with
the STM images\cite{si-highindex}) to determine the configuration
of all high-index Si surfaces, as long as the HOEP potential
remains satisfactory for all orientations to be investigated.
Furthermore, with certain modifications related to the
implementation of empirical potentials for systems with {\em two
atomic species}, the PTMC method could help bring important
advances in terms of finding the thermodynamically stable
intermixing composition of various nanostructures obtained by
heteroepitaxial deposition of thin films on silicon substrates.
Though such studies have recently been reported,\cite{landau-sige}
only the intermixing at a given atomic bonding topology has been
investigated. The interplay between reconstruction and intermixing
is another challenging and important problem that can be tackled
via PTMC simulations. Lastly, the method presented in this article
may also be used for studying the decomposition of unstable
orientations into nanofacets, as well as for predicting the
thermodynamics of surfaces in the presence of adsorbates or
applied strain.

{{\bf Acknowledgements.}} We gratefully acknowledge funding from
National Science Foundation through the Brown MRSEC program
(DMR-0079964), and Grants No. CHE-0095053 and CHE-0131114. The
simulations were performed at the Center for Advanced Scientific
Computation and Visualization at Brown University. We thank
Professor J.D. Doll for generous support, Professor M.C.
Tringides for useful discussions on the structure of high-index
silicon surfaces, and Professor S.J. Singer for valuable comments on
the manuscript.

\end{document}